

\input definit.tex
\magnification=1095
\baselineskip=18.0truept
\noindent \overfullrule=0pt
\par
\centerline{{\bf FLORY THEORY REVISITED}}
\medskip
\centerline{{\bf by}}
\smallskip
\centerline{{\bf H. Orland}\footnote {$ ^\ast $}{Also at Groupe de Physique
Statistique, Universit\'e de Cergy-Pontoise, 95806 Cergy-Pontoise
Cedex, France}}
\centerline{{\sl Service de Physique Th\'eorique\/}\footnote {$ ^{\dagger}
$}{Laboratoire de la Direction des Sciences de la Mati\`ere du Commissariat
\`a l'Energie Atomique}}
\centerline{{\sl Centre d'Etudes de Saclay, 91191 Gif-sur-Yvette Cedex,
France\/}}
\skiplines 4
\noindent{\bf ABSTRACT} \par
\skiplines 2
The Flory theory for a single polymer chain is derived
as the lowest order of a cumulant expansion. In this approach, the
full original Flory free energy ( including the logarithmic term ),
is recovered.
The prefactors of the elastic and repulsive energy are
calculated from the microscopic parameters. The method can be
applied to other types of monomer-monomer interactions, and
the case of a single chain in a bad solvent is discussed .
The method is easily
generalized to many chain systems (polymers in solutions), yielding
the usual crossovers with chain concentration.
Finally, this method is suitable for a systematic expansion around the Flory
theory. The corrections to Flory theory consist of
extensive terms ( proportional to the number $N$ of monomers ) and
powers of $N^{2-\nu d}$ . These last terms
diverge in the thermodynamic limit, but less rapidly than the usual Fixman
expansion in $N^{2- d/2}$. \par
\noindent \vglue 4truecm \par
\noindent Submitted for publication to \par
\noindent \lq\lq J. Phys. I \rq\rq\ \hfill{Saclay, SPhT/93-xxx} \par
\vfill\eject
\centerline{{\bf LA THEORIE DE FLORY REVISITEE}}
\vskip 2truecm
\noindent{\bf RESUME} \par
\skiplines 2
La th\'eorie de Flory pour une cha\^\i ne polym\'erique est obtenue
comme l'ordre dominant d'un d\'eveloppement en cumulants. Dans cette
approche, l'\'energie libre originale de Flory ( y compris le terme
logarithmique ) est obtenue. Les pr\'efacteurs des \'energies libres
\'elastique et r\'epulsive sont d\'eriv\'es \`a partir des
param\`etres microscopiques. La m\'ethode peut \^etre appliqu\'ee \`a
d'autres types d'interactions entre monom\`eres, et on discute le cas
d'une cha\^\i ne en mauvais solvant. La m\'ethode peut \^etre
g\'en\'eralis\'ee au cas de plusieurs cha\^\i nes ( solutions de
polym\`eres), et on en d\'eduit les changements de comportement en
fonction de la concentration en cha\^\i nes. Finalement, la m\'ethode
permet un d\'eveloppement syst\'ematique autour de la th\'eorie de
Flory. Les corrections \`a la th\'eorie de Flory comportent des termes
extensifs ( proportionnels au nombre $N$ de monom\`eres ) et des
puissances de $N^{2-\nu d}$. Ces termes divergent \`a la limite
thermodynamique, mais moins vite que le d\'eveloppement de Fixman, en
puissances de $N^{2- d/2}$. \par
\vfill\eject
\noindent{\bf I. INTRODUCTION} \par
\smallskip
The size of a polymer in a solvent is characterized by an
exponent $ \nu , $ which relates the radius of gyration $ R $ of the chain
to the degree of polymerization $ N $ (number of monomers) through
the formula$ ^{(1,2,3)} $
$$ R \sim  N^\nu \eqno (1) $$
\par
In an ideal $ \Theta $ solvent, chains are Brownian with $ \nu =1/2. $
In a good solvent, chains are swollen, and the index $ \nu $ is larger
than 1/2. \par
Flory$ ^{(1)} $ has devised a simple argument to compute the
exponent $ \nu $ of a chain in a good solvent, which gives amazingly
good agreement for all dimensions $ d. $ The argument, as presented
in ref.(2) goes as
follows: consider a chain of $ N $ monomers of length $ a, $ with local
repulsive interaction characterized by an excluded volume
parameter $ v. $ The swollen chain will acquire a radius of
gyration $ R; $ the monomer concentration is $ c=N/R^d, $ and the
repulsive energy is thus
$$ \matrix{\displaystyle E_{ {\rm rep}} & \displaystyle = {1 \over 2} v\ c^2\
R^d \hfill \cr\displaystyle  & \displaystyle = {1 \over 2} v {N^2 \over R^d}
\hfill \cr} \eqno (2a) $$
The elastic energy of the chain is given by:
$$ E_{ {\rm el}} = T {R^2 \over Na^2} \eqno (2b) $$
up to a multiplicative constant ( $T$ is the temperature). \par
The total free energy is:
$$ F = T {R^2 \over Na^2} + {v \over 2} {N^2 \over R^d} \eqno (3) $$
and minimization with respect to $ R $ yields the celebrated Flory
exponent:
$$ \nu_ F = {3 \over d+2} \eqno (4) $$
which is exact for $ d=1,2,4, $ and almost exact for $ d=3 $ (see
ref.(3); $ \nu_ F = 0.6 $ whereas best numerical estimates give $ \nu  =
0.588). $ \par
Note that this simplified derivation of the Flory theory misses a
logarithmic term in the free energy, which is present in the original
derivation of Flory (ref.(1) ).
In any case, the derivations of the Flory exponent are quite
empirical, and any attempt to go beyond it, by improving on any
of the two terms of (2a) or (2b) turns out to ruin the argument
(see ref.(2) for a discussion of that point). Moreover, the free
energy increases like $ N^{{4-d \over d+2}} $ instead of increasing like $ N $
as it
should in the asymptotic limit of large $ N $ (the free energy is not
extensive). \par
Des Cloizeaux has proposed a Gaussian variational
procedure to tackle this problem$ ^{(4)}, $ but the outcome was $ \nu_ c = {2
\over d}, $
by far too large a result. Also, Edwards has proposed a
self-consistent method$ ^{(5)} $ which yielded the Flory formula, but,
as noted by des Cloizeaux$ ^{(3)}, $ this approximation yields
unrealistic chain configurations. \par
In this paper, we propose a cumulant expansion,
which to lowest order generates a Flory type free energy from the
microscopic model, and which can be easily generalized
to the study of more complex systems (polyelectrolytes, chains in bad
solvents,
 membranes
and interfaces, solutions of polymers, etc...). This
method allows for a systematic expansion around Flory theory. \par
\medskip
\noindent{\bf II. ONE CHAIN SYSTEM} \par
\smallskip
We start from the Edwards continuous representation of a
chain:
$$ Z = \int^{ }_{\vec  r(N)=\vec  r(0)}{\cal D}\vec  r(s)\ \delta \left({1
\over N} \int^ N_0 {\rm d} s\ \vec  r(s) \right) {\rm exp} \left(-{1 \over
2} \int^ N_0 {\rm d} s\ \vdot  r^2-{1 \over 2} \int^ N_0 {\rm d} s {\rm d}
s^{\prime} v \left(\vec  r(s)-\vec  r \left(s^{\prime} \right) \right) \right)
\eqno (5) $$
where $ v \left(\vec  r-\vec  r^{\ \prime} \right) $ is the interaction
between a monomer at point $ \vec  r $ and
a monomer at $ \vec  r^{\ \prime}$ and units have been chosen so that
the persistence length $a$ is equal to 1.
\par
The first $ \delta $-function is used to constrain the center of
mass of the chain at the origin. Furthermore, to simplify
calculations, we have assumed that the chain is closed, i.e. $ \vec
r(N)=\vec  r(0). $
As will be seen later, this constraint is unessential, as far as
the method is concerned. \par
Performing a Gaussian transform on (5), and denoting by $ v^{-1}(r) $
the inverse of the kernel $ v \left(\vec  r \right), $ we have $ ^{(6)} $:
$$ Z = \int^{ }_{ }{\cal D}\phi \left(\vec  r \right) {\rm exp} \left(-{1
\over 2} \int^{ }_{ } {\rm d}\vec  r {\rm d}\vec  r^{\ \prime} \phi
\left(\vec  r \right)v^{-1} \left(\vec  r-\vec  r^{\ \prime} \right)\phi
\left(\vec  r^{\ \prime} \right) \right)Z_\phi \eqno (6a) $$
where
$$ Z_\phi  = \int^{ }_{\vec  r(N)=\vec  r(0)}{\cal D}\vec  r(s)\ \delta
\left({1 \over N} \int^ N_0 {\rm d} s\ \vec  r(s) \right) {\rm exp} \left(-{1
\over 2} \int^ N_0 {\rm d} s\ \vdot  r^2-i \int^ N_0 {\rm d} s\ \phi
\left(\vec  r(s) \right) \right) \eqno (6b) $$
\par
Using the identity:
$$
1 = \int_0^\infty {\rm d} Q\  \delta \left(Q-{1 \over N} \int^ N_0 {\rm d} s\
\vec
r^2(s) \right) \eqno (7a)
$$
where $Q$ is the square of the radius of gyration, we can rewrite (6b) as:
$$ Z_\phi  = \int^{ \infty}_ 0 {\rm d} Q\ Z_\phi( Q) \eqno (7b) $$
where
$$ \matrix{\displaystyle Z_\phi( Q) & \displaystyle = \int^{ }_{\vec
r(N)=\vec  r(0)}{\cal D}\vec  r(s)\ \delta \left({1 \over N} \int^ N_0 {\rm d}
s\ \vec  r(s) \right)\delta \left(Q-{1 \over N} \int^ N_0 {\rm d} s\ \vec
r^2(s) \right) \cr\displaystyle  & \displaystyle \times  {\rm exp} \left(-{1
\over 2} \int^ N_0 {\rm d} s\ \vdot  r^2-i \int^ N_0 {\rm d} s\ \phi( r(s))
\right) \cr} \eqno (7c) $$
\par
Defining $ \langle ...\rangle $ as the average with respect to the partition
function $ Z_0(Q) = Z_{\phi =0}(Q), $ we can
write
$$ \matrix{\displaystyle{ Z_\phi( Q) \over Z_0(Q)} & \displaystyle =
\left\langle {\rm exp} \left(-i \int^ N_0 {\rm d} s\ \phi( r(s)) \right)
\right\rangle \hfill \cr\displaystyle  & \displaystyle = \left\langle {\rm
exp} \left(-i \int^{ }_{ } {\rm d}\vec  r\ \phi \left(\vec  r \right)\rho
\left(\vec  r \right) \right) \right\rangle \hfill \cr} \eqno (8) $$
where $ \rho \left(\vec  r \right) = \int^ N_0 {\rm d} s\ \delta \left(\vec
r-\vec  r(s) \right). $ \par
We use the cumulant expansion:
$$ \left\langle {\rm e}^{-A} \right\rangle  = {\rm exp} \left(-\langle
A\rangle +{1 \over 2} \left( \left\langle A^2 \right\rangle -\langle
A\rangle^ 2 \right) - {1 \over 6} \left( \left\langle A^3 \right\rangle -3
\left\langle A^2 \right\rangle\langle A\rangle +2\langle A\rangle^ 3
\right)+... \right) \eqno (9) $$
To lowest order, we have:
$$ Z_\phi( Q) \simeq  Z_0(Q)\ {\rm exp} \left(-i \int^{ }_{ } {\rm d}\vec  r\
\phi \left(\vec  r \right)\ \rho_ Q \left(\vec  r \right) \right) \eqno (10a)
$$
where
$$ Z_0(Q) = \int^{ }_{\vec  r(N)=\vec  r(0)}{\cal D}r(s)\ \delta \left({1
\over N} \int^ N_0 {\rm d} s\ \vec  r(s) \right)\delta \left(Q-{1 \over N}
\int^ N_0 {\rm d} s\ \vec  r^2(s) \right) {\rm exp} \left(-{1 \over 2}
\int^ N_0 {\rm d} s\ \vdot  r^2(s) \right) \eqno (10b) $$
and
$$ \matrix{\displaystyle \rho_ Q(\vec r) & \displaystyle = \langle
\rho( \vec r)\rangle
\hfill \cr\displaystyle  & \displaystyle = {1 \over Z_0(Q)} \int^ N_0 {\rm d}
s \int^{ }_{\vec  r(N)=\vec  r(0)}{\cal D}\vec  r(s)\ \delta \left({1 \over N}
\int^ N_0 {\rm d} s\ \vec  r(s) \right)\delta \left(Q-{1 \over N} \int^ N_0
{\rm d} s\ \vec  r^2(s) \right) \hfill \cr\displaystyle  & \displaystyle \ \ \
\ \ \ \ \ \ \ \ \ \times  {\rm exp} \left(-{1 \over 2} \int^ N_0 {\rm d} s\
\vdot  r^2(s) \right)\delta \left(\vec  r-\vec  r(s) \right) \hfill \cr}
\eqno (10c) $$
$ Z_0(Q) $ and $ \rho_ Q(\vec r) $ are respectively the partition function and
monomer concentration for a Brownian chain
with center of mass constrained at the origin and radius of gyration square
constrained to $ Q. $ \par
Finally, performing the $ \{ \phi\} $ Gaussian integral of (6a), we obtain:
$$ Z \simeq  Z_F = \int^{ \infty}_ 0 {\rm d} Q\ Z_0(Q)\ {\rm exp} \left(-{1
\over 2} \int^{ }_{ } {\rm d}\vec  r {\rm d}\vec  r^{\ \prime} \rho_ Q
\left(\vec  r \right)v \left(\vec  r-\vec  r^{\ \prime} \right)\rho_ Q
\left(\vec  r^{\ \prime} \right) \right) \eqno (11) $$
The $ F $ index in (11) stands for Flory, since, as we shall see, $ Z_F $ is
just the Flory partition
function. \par
A straightforward calculation of (10c) and (10d) yields
( see the appendix ), for $ Q/N \longrightarrow \infty$ (swollen chain) :
$$ Z_0(Q) \simeq  { {(2\pi)}^{2d} \over (d-1)!} \left( {Q \over N}
\right)^{(d-1)}
\exp
\left(-{2\pi^2 Q \over N} \right) \eqno (12a) $$
and
$$ \rho_ Q \left(\vec  r \right) \simeq  N \left({d \over 2\pi Q}
\right)^{d/2} {\rm exp} \left(- {d \over 2Q} \vec  r^2 \right) \eqno (12b) $$
\par
In the thermodynamic limit, $ N \longrightarrow +\infty , $ and (11) can be
evaluated by applying the saddle-point
method on $ Q. $ Assuming a standard contact interaction:
$$ v \left(\vec  r \right) = v\ \delta \left(\vec  r \right) \eqno (13) $$
where $ v $ is the excluded volume parameter, the function to minimize is:
$$ \beta F_F = -(d-1)\ {\ln} \left({Q \over N} \right) + 2\pi^ 2 {Q \over
N} + {v \over 2} \left({d \over 4\pi} \right)^{d/2} {N^2 \over Q^{d/2}}
\eqno (14a) $$
where $\beta$ is the inverse temperature,
and the radius of gyration $ R_G= \sqrt{ Q} $ satisfies the minimization
equation:
$$ - {2(d-1) \over R_G} + 4\pi^ 2 {R_G \over N} - {vd \over 2} \left({d \over
4\pi} \right)^{d/2} {N^2 \over R_G^{d+1}} = 0 \eqno (14b) $$
\par
We recognize in equations (14) the Flory free energy, with calculated
prefactors, including
the logarithmic term, present in the original Flory theory. The role of this
term is just to ensure a
crossover from a Brownian regime in $ d>4 $ to a swollen regime in $ d\leq 4.
$ Indeed, to solve (14b), we must
balance two terms of opposite sign, and check that the third one is
negligeable with respect to them. \par
Balancing the $ 2^{ {\rm nd}} $ and $ 3^{ {\rm rd}} $ term in (14b) yields the
Flory result, $ \nu_ F = {3 \over d+2}. $ The consistency
requirement: $ {1 \over R_G} \ll  {R_G \over N} $ is satisfied provided that $
d\leq 4. $ \par
Balancing the $ 1^{ {\rm st}} $ and $ 2^{ {\rm nd}} $ term in (14b) yields the
Brownian exponent $ \nu =1/2, $ and the
consistency check $ {N^2 \over R_G^{d+1}} \ll  {R_G \over N} $ is satisfied for
$ d\geq 4. $ \par
Thus the logarithmic term indeed, enforces the crossover between the 2
regimes. \par
The method can be applied to any type of two-body interaction $ v \left(\vec  r
\right) $ and denoting by $ v (\vec  k ) $ its Fourier transform, the
Flory free energy becomes:
$$
 \beta F_F = -(d-1)\ {\ln} \left({Q \over N} \right) + 2\pi^ 2 {Q \over
N} - {\beta \over 2} \int {\rm d} {\vec k} v (\vec  k )
\exp \left( -Q {{\vec k}^2 \over d} \right)
\eqno (14c)
$$
Application of this method to the case of polyelectrolytes, where the two-body
interaction is the Coulomb potential $ v \left(\vec  r
\right) = { 1 \over {r^{(d-2)}}} $ , yields the well-known Flory
exponent $ ^{(7)} $ $ \nu = {3 \over d} $ . This is known to be
incorrect for dimensions larger than $3$ , where the exact exponent
is $ \nu = {2 \over d-2 }$ ( see ref(7) ).

Finally, let us show how this method works for a chain in a
poor solvent. In that case, in addition to the second virial
coefficient $v$ which is attractive, one has to introduce the third virial
coefficient
$w$ which is repulsive.
 The partition function of the chain reads:
$$
Z = \int^{ }_{\vec  r(N)=\vec  r(0)}{\cal D}\vec  r(s)\ \delta \left({1
\over N} \int^ N_0 {\rm d} s\ \vec  r(s) \right)
$$
\nobreak
$$
 {\rm exp} \left(-{1 \over
2} \int^ N_0 {\rm d} s\ \vdot  r^2+{v \over 2} \int^ N_0 {\rm d} s {\rm d}
s^{\prime} \delta \left(\vec  r(s)-\vec  r \left(s^{\prime} \right)
\right)
- {w \over 6}  \int^ N_0 {\rm d} s {\rm d} s^{\prime}{\rm d} s^{\prime \prime}
\delta \left(\vec  r(s)-\vec  r \left(s^{\prime} \right) \right)
\delta \left(\vec  r(s)-\vec  r \left(s^{\prime \prime} \right)
\right)
\right)
\eqno (15a)
$$
In the present case, the alternative to the Gaussian transform is to
constrain the monomer density $ \rho ( \vec r ) $
as a new integration variable through the identity:
$$
1= \int {\cal D} \phi ( \vec r ){\cal D} \rho ( \vec r )
\exp \left ( i \int {\rm d} \vec r \phi ( \vec r ) \rho ( \vec r )
-i \int^ N_0 {\rm d} s\ \phi \left(\vec  r(s) \right) \right)
\eqno (15b)
$$
Inserting identity (15b) in equation (15a) yields:
$$
Z = \int {\cal D} \phi ( \vec r ){\cal D} \rho ( \vec r )
\exp \left ( i \int {\rm d} \vec r \phi ( \vec r ) \rho ( \vec r )
+ {v \over 2} \int {\rm d} \vec r \rho^2 (\vec r)
- {w \over 6} \int {\rm d} \vec r \rho^3 (\vec r) \right )
Z_\phi
\eqno (15c)
$$
where $Z_\phi$ is defined by eq.(6b). Using identity (7a) in the above equation
together with the lowest order cumulant ( see eq.(10a))
and integrating over the fields $\phi$ and $\rho$ yields the expression:
$$
Z \simeq \int_0^\infty {\rm d} Q Z_0(Q)\ \exp \left (
+ {v \over 2} \int {\rm d} \vec r \rho_Q^2 (\vec r)
- {w \over 6} \int {\rm d} \vec r \rho_Q^3 (\vec r) \right )
\eqno (16)
$$
where $ Z_0(Q)$ and $\rho_ Q \left(\vec  r \right)$ are defined in eq.(10 b,c).

In the case $ v < 0 $, we are back to the previous situation, and the chain
is swollen. The free energy reads:
$$
 \beta F_F = -(d-1)\ {\ln} \left({Q \over N} \right) + 2\pi^ 2 {Q \over
N} - {v \over 2} \left({d \over 4\pi} \right)^{d/2} {N^2 \over
Q^{d/2}}
+ {w \over 6} \left ({d \over 18\pi} \right)^{d} {N^3 \over
Q^{d}}
\eqno (17a)
$$

At the $\Theta$ point, $ v = 0 $, and the free energy is still given
by the above expression with $ v = 0 $ .
This regime corresponds to an exponent $\nu= {2 \over {d+1}}$.
In $ d=3 $, as is well known, this yields a Brownian exponent $ \nu=
1/2 $. Note that , in that case, since the free energy of equation
(17a) is finite, one cannot evaluate
the integral (16) by the saddle-point method.
In that case, quadratic corrections must be
included to compute the free energy.

In the collapsed regime, $ Q/N \longrightarrow 0 $ and
$ Z_0(Q)$ is no more given by (12a), although
the asymptotic form of $\rho_ Q \left(\vec  r \right)$ is given by
(12b).
We show in the appendix that the correct form for $ Z_0(Q)$
in dimension $ d=3 $ is:
$$
Z_0(Q) = {3 \over {16 \sqrt(2\pi)}} \left ({ N \over Q} \right ) ^{5/2}
\exp(-{N \over {8 Q}})
\eqno (17b)
$$
yielding  a different Flory free energy:
$$
 \beta F_F = - \ln \left ({3 \over {16 \sqrt(2\pi)}} \left ({ N \over Q} \right
) ^{5/2}
\right )
+ {N \over {8 Q}}
- {v \over 2} \left({3 \over 4\pi} \right)^{3/2} {N^2 \over
Q^{3/2}}
+ {w \over 6} \left ({1 \over 6\pi} \right)^{3} {N^3 \over
Q^{3}}
\eqno (17c)
$$
In this regime, we recover the usual exponent  $\nu= {1 \over 3}$.

The method described above, can be applied to any type of interaction
$ v \left(\vec  r\right), $ and can be easily generalized
to the case of membranes or interfaces $^{(8)}$. \par
\medskip
\noindent{\bf III. PAIR CORRELATION FUNCTION} \par
\smallskip
At this level of approximation, it is easy to compute the
corresponding pair correlation function $ g ( \vec r )$ .
Using the following definition:
$$
g(\vec r) = {1 \over N} \int_0^{N} {\rm d}s \int_0^{N} {\rm d}s'
\ \langle \delta (\vec r(s))
\delta (\vec r - \vec r(s')) \rangle
\eqno (18a)
$$
or
$$
g(\vec q) = {1 \over N} \int_0^{N} {\rm d}s \int_0^{N} {\rm d}s' \
\langle \exp( i \vec q ( \vec r(s) - \vec r(s'))) \rangle
\eqno (18b)
$$
in Fourier space.
Using the lowest order cumulant expansion (10a), we show in the appendix
that the pair correlation function is given by:
$$
g(\vec q) = N (d-1)! \int_{0}^{\pi} {{\rm d}\theta
\over {\pi}} \left( {1 \over Q (1-\cos\theta) {\vec q}^2 } \right )
^{d-1 \over 2 } J_{d-1} \left ( 2 \sqrt{Q (1-\cos\theta) {\vec q}^2 }\right )
$$
$$
\times \exp \left (- {N {\vec q}^2 \over 4\pi^2 } ( 1 + {\cos\theta \over 2} -
(\pi-\theta) \sin\theta)\right)
\eqno (19a)
$$
where the square radius of gyration $Q$ is given by its saddle-point value as
computed in the previous section, and $J_{d-1}$ denotes the Bessel
function of integer order $(d-1)$ .

At small $\vec q$, i.e. large distances
$ {\vec q}^2 Q <<1 $ , the function reduces to:
$$
g(\vec q) = N \int_0^{\pi} {{\rm d}\theta \over \pi}
\exp( - Q (1-\cos\theta) {\vec q}^2 /d)
\eqno (19b)
$$
whereas at smaller distances $ {\vec q}^2 Q > 1 $ , a simple scaling
argument shows that:
$$
g(\vec q) \simeq { 1 \over q^{1/\nu} } \simeq { 1 \over q^{5/3} }
\eqno (19c)
$$
in three dimensions. This result agrees with the prediction of Edwards$^{(5)}$.
\medskip
\noindent{\bf IV. MANY CHAINS SYSTEMS} \par
\smallskip
Consider now a system of $ {\cal N} $ polymer chains of length $ N $
(polydispersity effects can be trivially
included). The partition function of the system reads:
$$ Z = \int^{ }_{\vec  r_i(N)=\vec  r_i(0)} \prod^{{\cal N}}_{i=1}{\cal
D}\vec  r_i(s)\ {\rm exp} \left(-{1 \over 2} \sum^{{\cal N}}_{i=1} \int^
N_0 {\rm d} s\ \vdot  r^2_i-{v \over 2} \sum^{{\cal N}}_{i,j=1} \int^ N_0
{\rm d} s {\rm d} s^{\prime} \delta \left(\vec  r_i(s)-\vec  r_j(s) \right)
\right) \eqno (20) $$
\par
Introducing a center of mass coordinate $ \vec  R_i $ and a square radius of
gyration $ Q_i $ for each chain,
the same procedure as in II) yields the following expression:
$$ \matrix{\displaystyle Z\simeq Z_F & \displaystyle = \int^{ }_{ }
\prod^{{\cal N}}_{i=1} {\rm d}\vec  R_i {\rm d} Q_i\ {\rm exp} \left(
\sum^{{\cal N}}_{i=1} \left( {(d-1)\ \ln} \left({Q_i \over N}
\right)-2\pi^ 2{Q_i \over N} \right) \right. \hfill \cr\displaystyle  &
\displaystyle \left.- v {N^2 \over 2} \sum^{ }_{ i,j} \left({d \over 2\pi
\left(Q_i+Q_j \right)} \right)^{d/2} {\rm exp} \left(-{d \over 2} {
\left(\vec  R_i-\vec  R_j \right)^2 \over Q_i+Q_j} \right) \right) \hfill
\cr} \eqno (21a) $$
\par
Each polymer chain is represented by a coordinate $ \vec  R_i $ (center of
mass) and a radius of gyration
square $ Q_i. $ The determination of the size $ Q_i $ will be done by a
saddle-point method, and it is natural
(for monodisperse systems) to assume that all sizes are equal: $ Q_i=Q. $ We
obtain:
$$ \matrix{\displaystyle Z_F & \displaystyle = {\rm exp} \left({\cal N}
\left[{(d-1)\ \ln} \left({Q \over N} \right)-2\pi^ 2 {Q \over N} \right]
\right) \hfill \cr\displaystyle  & \displaystyle \times  \int^{ }_{ }
\prod^{{\cal N}}_{i=1} {\rm d}\vec  R_i\ {\rm exp} \left(-{vN^2 \over 2}
\left({d \over 4\pi Q} \right)^{d/2} \sum^{{\cal N}}_{i,j=1} {\rm exp} \left(-
{d \over 4Q} \left(\vec  R_i-\vec  R_j \right)^2 \right) \right) \hfill \cr}
\eqno (21b) $$
\par
Thus, the partition function contains a Flory-like term, characteristic of
each individual
chain properties, and a partition function representing a liquid of
interacting chains with a smoothed
interaction:
$$ V \left(\vec  R \right) = vN^2 T \left({d \over 4\pi Q} \right)^{d/2} {\rm
exp} \left(- {d \over 4Q} \vec  R^2 \right) \eqno (22) $$
\par
The range of the interaction is the size of the chains $ \sqrt{ 2Q}. $ Once $
Z_F $ is calculated, the radius of gyration $R_G$
is to be determined by minimizing the total free energy of the system
with respect to $Q$.

Defining the monomer concentration $ c = N{\cal N} /\Omega $ ( where
$\Omega $ is the volume ) , and
the overlap threshold concentration
$ c^* = { N \over R_G^d} \simeq N ^{ (1 - \nu d) }$
( where $R_G = \sqrt Q $ is the radius of gyration of a chain, and $\nu$
is the Flory exponent $ \nu = { 3 \over (d+2) } $ )
, the various regimes of chain
concentration can easily be recovered.  \par
\smallskip
i) {\sl Dilute regime\/} $ c << c^* $: it is the gas type regime, where the
density of
chains is such that the typical
interchain distance is much larger than the chain size $ R_G $ . In that
case, we have a gas of weakly
interacting single chains.
The partition function can be evaluated by using the virial expansion $^{(9)}$,
and the osmotic pressure reads:
$$
\Pi = T { c \over N} (1 + A_2 {c \over N} + ...)
\eqno (23a)
$$
where $c/N = {\cal N} / \Omega $ is the polymer concentration, and $A_2$ is
the second virial coefficient calculated with (22):
$$
A_2 = - {1 \over 2} \int {\rm d} {\vec R} \left( \exp(-\beta V(\vec R)) - 1
\right)
\eqno (23b)
$$
A simple evaluation of this integral yields:
$$
A_2 \simeq  R_G^d\ (\ln N)^{d/2}
\eqno (23c)
$$
Note the logarithmic term in the virial coefficient $A_2$,
absent from standard calculations.
The total free energy per polymer reads:
$$
 F / {\cal N} =   F_F + c/N A_2 T
\eqno (23d)
$$
where $F_F$ is the single chain Flory free energy as given by (14a).
Minimization of (23d) with respect to $Q$ yields the usual Flory exponent.
\par
\smallskip
iii) {\sl Melt\/} $ c^* << c \simeq 1/a^d $ ( where $a$ is the monomer size )
: when interchain distances become smaller than the radius $
R_G, $ the chains overlap
strongly, and the monomer concentration is constant, with small fluctuations.
Using a Gaussian transform, equation (21b) can be written as:
$$ \matrix{\displaystyle Z_F & \displaystyle = {\rm exp} \left({\cal N}
\left[{(d-1)\ \ln} \left({Q \over N} \right)-2\pi^ 2 {Q \over N} \right]
\right) \hfill \cr
\displaystyle
& \displaystyle \times  \int
{\cal D}\phi(\vec R) \ {\rm exp} \left(-{\beta \over 2}
\int {\rm d} \vec R \int {\rm d} \vec R' \phi(\vec R) V^{-1}(\vec R - \vec R')
\phi(\vec R')
+ {\cal N} \ln ( \int {\rm d}\vec R \exp( i \beta \phi(\vec R))) \right )
\hfill \cr}
\eqno (24a) $$
where $ V^{-1} ( \vec R - \vec R' ) $ is the inverse kernel of $ V ( \vec R
- \vec R' ) $.
\par This functional integral can be evaluated by the saddle-point method.
Assuming a uniform monomer concentration is equivalent to assume a constant
field $ \phi ( \vec R) = \phi_0 $. A simple calculation shows that the
saddle-point is given by:
$$
 \beta \phi_0 = i N c v^2
\eqno (24b)
$$
so that the free energy becomes:
$$
\beta F_F \simeq  -{\cal N} \left( (d-1) \ln {Q \over N} - 2\pi^2 {Q
\over N} + \ln \Omega \right )
+ {v (N{\cal N})^2 \over 2 \Omega } \eqno (24c)
$$
Minimization of (19c) with respect to $Q$ yields the standard Brownian
exponent $\nu = 1/2$ , and the osmotic pressure is given by:
$\Pi \beta = {v \over 2} c^2$
\par
The interchain interactions screen out completely, and the
 chains become Brownian,
as is well
known$ ^{(10)}. $  A systematic expansion around the mean-field
$\phi_0$ can easily be done, and the lowest order turns out to be
identical to the usual RPA $ ^{(6)} $ .
\smallskip
ii) {\sl Semi-dilute regime\/}  $ c^* << c << 1/a^d $
: it is the liquid type regime, which requires
more involved
calculations of the liquid-like partition function$ ^{(9)} $.
A careful analysis of the generic term of the virial expansion of
(16b) leads to the scaling form proposed by Des Cloizeaux $ ^{(11)} $:
$$
\beta \Pi = c/N f \left( {c \over N} R_G^d \right)
\eqno(25a)
$$
which under a scaling hypothesis leads to the well-known behaviour:
$$
\beta \Pi = c^{9 \over 4}
\eqno (25b)
$$
in dimension $d=3$
\par
\medskip
\noindent{\bf V. CORRECTIONS TO THE FLORY THEORY} \par
\smallskip
One great advantage of this method is that it allows to compute corrections
to the Flory
theory. Indeed, using eq.(9) to second order, we have:
$$ Z_\phi( Q)\simeq Z_0(Q)\ {\rm exp} \left(-i \int^{ }_{ } {\rm d}\vec  r\
\phi \left(\vec  r \right)\rho_ Q \left(\vec  r \right) - {\lambda \over 2}
\int^{
}_{ } {\rm d}\vec  r {\rm d}\vec  r^{\ \prime} \phi \left(\vec  r \right)G_Q
\left(\vec  r,\vec  r^{\ \prime} \right)\phi \left(\vec  r^{\ \prime} \right)
\right) \eqno (26a) $$
where $ G_Q \left(\vec  r,\vec  r^{\ \prime} \right) $ is the
connected Debye function$
^{(12)} $ of a Brownian chain with constrained square radius of gyration
$ Q: $
$$ G_Q \left(r,r^{\prime} \right) = \int^ N_0 {\rm d} s {\rm d} s^{\prime}
\left( \left\langle \delta \left(\vec  r-\vec  r(s) \right)\ \delta
\left(\vec  r^{\ \prime} -\vec  r \left(s^{\prime} \right) \right)
\right\rangle - \left\langle \delta \left(\vec  r-\vec  r(s) \right)
\right\rangle \left\langle \delta \left(\vec  r^{\ \prime} -\vec  r
\left(s^{\prime} \right) \right) \right\rangle \right) \eqno (26b) $$
and is given in momentum space by (see appendix):
$$ G_Q \left(\vec k,\vec k^{\prime} \right)
= {N^2 \over 2\pi}
\int^{2\pi}_0 {\rm d} u
\ {\rm exp} \left(-{Q \over 2d} (\vec k^2 + \vec k'^2) \right)
\left( \exp\left(-({Q\over d}\cos u -{N \over 4 \pi}\sin u)\vec k \vec k'
\right) -1 \right)
\eqno(26c)
$$
\par
The parameter $\lambda$ in (26a) is just a remainder to keep trace of
the orders of the expansion.
\par
Integration over the $ \phi $ variables yields:
$$ \matrix{\displaystyle Z & \displaystyle = \int^{ \infty}_ 0 {\rm d} Q\
{\rm exp} \left(- \left(- (d-1)\ {\rm \ln} \left({Q \over N}
\right)+2\pi^ 2 {Q \over N} \right)-{1 \over 2} {\rm Tr\ \ln}
\left(\delta \left(\vec  r-\vec  r^{\ \prime} \right)+
\lambda v\ G_Q \left(\vec
r,\vec  r^{\ \prime} \right) \right) \right. \cr\displaystyle  &
\displaystyle \left.-{v \over 2} \int^{ }_{ } {\rm d}\vec  r {\rm d}\vec  r^{\
\prime} \rho_ Q \left(\vec  r \right) \left(\delta \left(r-r^{\prime}
\right)+ \lambda vG_Q \left(\vec  r,\vec  r^{\ \prime} \right)
\right)^{-1}\rho_ Q
\left(\vec  r^{\ \prime} \right) \right) \cr} \eqno (27a) $$
\par
Expanding to lowest order in $ \lambda, $ we obtain:
$$ \beta F_1 = \beta F_F +
\lambda \left({v \over 2} \int^{ }_{ } {\rm d}\vec  r\ G_Q
\left(\vec  r,\vec  r \right)-{v^2 \over 2} \int^{ }_{ } {\rm d}\vec  r {\rm
d}\vec  r^{\ \prime} \rho_ Q \left(\vec  r \right)G_Q \left(\vec  r,\vec  r^{\
\prime} \right)\rho_ Q \left(\vec  r^{\ \prime} \right)\right) \eqno (27b) $$
\par
The first correction term is thus of the form:
$$
\eqalign{
\int {\rm d}\vec r\ G_Q(\vec r,\vec r) &= \int {{\rm d} \vec k \over
(2\pi)^d } \ G_Q(\vec k, -\vec k) \cr
&={N^2 \over \pi}
\int {{\rm d} \vec k \over
(2\pi)^d }
\int^{\pi}_{2\pi a \over N }{\rm d} u
\ {\rm exp} (-{Q \over d} \vec k^2 )
\left( \exp\left(({Q\over d}\cos u -{N \over 4 \pi}\sin u)\vec k^2
\right) -1 \right) \cr
&= A N + B {N^2 \over Q^{d/2}} \cr
}
\eqno (28a)
$$
where $A$ and $B$ are finite numerical constants depending on
the monomer size $a$ and space dimension $d$, and
the natural cut-off ${2 \pi a \over N}$ has been introduced to
avoid ultra-violet divergences. The term linear in $N$ is the expected
extensive part of the free energy, which in the Flory expansion
appears thus as a correction term, and the second term is a correction
to the standard Flory term in the repulsive energy.

The second correction term is given by:
$$ \int{\rm d}\vec  r {\rm d}\vec  r^{\ \prime} \rho_ Q \left(\vec  r
\right)G_Q \left(\vec  r,\vec  r^{\ \prime} \right)\rho_ Q \left(\vec  r^{\
\prime} \right) = C \left({N^2 \over Q^{d/2}} \right)^2 \eqno (28b) $$
where $ C $ is a finite constant, depending only on the monomer
size $ a $ and the dimension $ d $ . This term scales like the
square of the Flory repulsive energy.

More generally, it can be proven that the expansion of the free energy
around the Flory
theory will generate extensive terms, as well as
powers of $ {N^2 \over Q^{d/2}} $ .
This is in contrast with the usual Fixman expansion $ ^{(13)} $
, which also contains extensive terms, but which is done in powers of $
N^{2-d/2} $. The Flory expansion is in terms of $ N^{2-\nu d} = N^{4-d
\over d+2 } $ .
\par
As is often the case when calculating corrections to mean field theories,
correction terms are much larger than
the mean field contribution, in the critical region $ (N \longrightarrow
+\infty ). $ \par
This allows the definition of a Ginzburg region$ ^{(14)}, $ i.e. a typical size
$ N_{ {\rm max}} $ such that for $ N < N_{ {\rm max}}, $ the
correction terms are small compared to the Flory free energy. The precise
value of $ N_{ {\rm max}} $ depends on all the
parameters of the problem, and we have not computed it explicitly, since it
is in fact a crossover size, and its precise value is not very illuminating.

However, it is clear that the Flory expansion will have a much larger
domain of validity than the Fixman expansion, since for any dimension
smaller than $4$, we have $ N^{4-d \over d+2} << N^{4-d \over 2} $ ,
and therefore,
this slower divergence of the Flory expansion might be the clue to its
success.
\medskip
\noindent{\bf VI. CONCLUSION} \par
\smallskip
We have shown how the original Flory theory of polymers can be derived
rigorously from a cumulant expansion.
We obtain a Flory like free energy, with the correct original
logarithmic term.
This method can be generalized to other types of
monomer interactions (e.g. bad solvent,
polyelectrolytes, etc...) and can be applied to solutions of polymers in a
straightforward manner. It can also
certainly be useful to other classes of problems (membranes, interfaces,
etc...). Finally, we show how this method can
generate a systematic expansion around the Flory theory. The calculations are
somewhat cumbersome, but they show that
the leading corrections diverge when $ N \longrightarrow +\infty  $, but
much less rapidly than the usual Fixman expansion.
\medskip
\noindent{\bf Acknowledgements} \par
\smallskip
The author wishes to thank J. Bascle, J.F. Joanny and T. Garel for useful
discussions, and
specially J. des Cloizeaux for a
critical reading of the manuscript. \par
\medskip
\vfill\eject
\centerline{{\bf REFERENCES}}
\medskip
\noindent \item {(1)}P. Flory, {\sl Principles of Polymer Chemistry\/},
Cornell
University Press, Ithaca, N.Y. (1971). \par
\smallskip
\noindent \item {(2)}P.G. de Gennes, {\sl Scaling Concepts in Polymer
Physics,\/}
Cornell University Press, Ithaca, N.Y. (1979). \par
\smallskip
\noindent \item {(3)}J. des Cloizeaux and G. Jannink, {\sl Polymers Solution,
their
Modelling and Structure, \/}Clarendon Press, (1990). \par
\smallskip
\noindent \item {(4)}J. des Cloizeaux, {\sl J. Physique\/} {\bf 31} (1970)
715. \par
\smallskip
\noindent \item {(5)}S.F. Edwards, {\sl Proc. Phys. Soc. London\/} {\bf 85}
(1965) 613. \par
\smallskip
\noindent \item {(6)}S.F. Edwards, {\sl Proc. Phys. Soc. London\/} {\bf 88}
(1966) 265. \par
\smallskip
\noindent \item {(7)}P. Pfeuty, R. Velasco and P.G. de Gennes,
{\sl J. Physique Lett.\/} {\bf 38L}
(1977) 5. \par
\smallskip
\noindent \item {(8)}H. Orland, to be published. \par
\smallskip
\noindent \item {(9)}J.P. Hansen and J.R. Mc Donald, {\sl Theory of Simple
Liquids,\/}
Academic Press, London (1976). \par
\smallskip
\noindent \item {(10)}P. Flory, {\sl J. Chem. Phys.\/} {\bf 17} (1949) 303.
\par
\smallskip
\noindent \item {(11)}J. des Cloizeaux, {\sl J. Physique\/} {\bf 36}
(1975) 281. \par
\smallskip
\noindent \item {(12)}P. Debye, {\sl J. Phys. Colloid Chem.\/} {\bf 51} (1947)
18; see also ref.(2). \par
\smallskip
\noindent \item {(13)}M. Fixman, {\sl J. Chem. Phys.\/} {\bf 23}
(1955) 1656; see also E. Teramoto,, {\sl Busseiron Kenkyu\/} {\bf 39}
(1951) 1. \par
\smallskip
\noindent \item {(14)}V.L. Ginzburg and L.D. Landau, {\sl JETP\/} {\bf 20}
(1950) 1064. \par
\smallskip
\noindent \item {(15)}I.S. Gradshteyn and I.M. Ryzhik,
{\sl Table of integrals, series, and products\/}, Academic Press ( New-York
and London )
(1965).  \par
\vfill\eject
\centerline{{\bf APPENDIX}}
\medskip
In this section, we show how to compute partition and correlation
functions for a Brownian chain with constrained center of mass, and
constrained radius of gyration.
Consider the partition function:
$$
Z_0(Q) = \int_{\vec  r(N)=\vec  r(0)}{\cal D}r(s)\ \delta \left({1
\over N} \int^ N_0 {\rm d} s\ \vec  r(s) \right)\delta \left(Q-{1 \over N}
\int^ N_0 {\rm d} s\ \vec  r^2(s) \right) {\rm exp} \left(-{1 \over 2}
\int^ N_0 {\rm d} s\ \vdot  r^2(s) \right)
\eqno (A1)
$$
In order to compute it, we expand the trajectories $\vec r(s)$ as
Fourier series:
$$
\eqalign{
\vec r(s) &= \sum_{n=-\infty}^{+\infty} e^{ i{2\pi \over N} s} \vec
r_n \cr
\vec r_n &= { 1\over N} \int_0^N {\rm d} s e^{ -i{2\pi \over N} s} \vec
r(s) \cr
}
\eqno(A2)
$$
The center of mass constraint implies that the Fourier component $\vec
r_0$ vanishes. The $\delta-$ function constraining the
radius of gyration can be represented by its Fourier integral, and
then the remaining Gaussian integral on $\vec r_n$
can be performed. After some
simple algebra, we obtain:
$$
Z_0(Q) = \int_{-\infty}^{+\infty} {{\rm d} z \over 2\pi}
\ e^{izQ} \prod_{n=1}^{+\infty} \left( {1 \over 1+{izN \over
2\pi^2 n^2}} \right)^d
\eqno(A3)
$$
This integral can be computed by the method of residues. The poles of
the function are given by:
$$
z_n = {2 i \pi^2 n^2 \over N}
\eqno(A4)
$$
and they are of order $d$ .

Using the analytic expression for the infinite product $^{(15)}$:
$$
\prod_{n=1}^{+\infty} (1 + {x^2 \over n^2 \pi^2}) = {\sinh x \over x}
\eqno(A5)
$$
we can write:
$$
Z_0(Q) = \int_{-\infty}^{+\infty} {{\rm d} z \over 2 N \pi} \ f(z)
\eqno(A6)
$$
with
$$
f(z)= e^{izq} \left( {\sqrt{iz \over 2} \over \sinh{\sqrt{iz \over 2}}}
\right)^d
\eqno(A7)
$$
where we have defined $q={Q \over N}$.
In the following, we will forget the $N$ factor appearing in the denominator
of (A6), since it is just a normalization constant. We have thus:
$$
Z_0(Q) = i \sum_{n=1}^{+\infty} {\rm res} (f,z_n)
\eqno(A8)
$$
where res denotes the residue of the function.

The exponential factor in the function $f$ implies terms of the form
$\exp(-2n^2\pi^2 q)$ in the sum (A8).

In the case of a swollen chain, the exponent $\nu$ is larger than
$1/2$, and thus, $ q \longrightarrow +\infty$ when $ N \longrightarrow
+\infty$. Therefore, in the sum (A8), only the pole $n=1$ will contribute,
all other poles being exponentially subdominant.
The sum (A8) reduces to:
$$
Z_0(Q) \simeq  { {(2\pi)}^{2d} \over (d-1)!} \left( {Q \over N} \right)^{(d-1)}
\exp
\left(-{2\pi^2 Q \over N} \right)
\eqno (A9) $$

In the case of a collapsed chain, the exponent $\nu$ is smaller than
$1/2$, and thus, $ q \longrightarrow 0$ when $ N \longrightarrow
+\infty$. Therefore, in the sum (A8), all poles contribute
equally, but the calculation of the residue is simplified by the fact
that $q$ is small. Since the calculation of the residue of a pole of
order $d$ involves the calculation of a
derivative of order $d$, we will specialize to dimension $d=3$.
A simple calculation of the residues yield the asymptotic formula:
$$
Z_0(Q) = - \sum_{n=1}^{+\infty} (-1)^n 6n^2\pi^2\ e^{-2n^2\pi^2 q}
\eqno(A10)
$$
for $ {Q\over N}\longrightarrow 0$.
Equation (A10) can be recast in a form which emphasizes its
ressemblance to the Jacobi elliptic Theta functions$^{(15)}$:
$$
Z_0(Q) = 3 {{\rm d} \over {\rm d}q}\  \sum_{n=1}^{+\infty}
\exp(-2n^2\pi^2 q + i n \pi)
\eqno(A11)
$$
Using the Poisson summation formula, we obtain:
$$
Z_0(Q) = {3 \over 2} {{\rm d} \over {\rm d}q}\
\sum_{n=-\infty}^{+\infty} e^{- {(2n+1)^2 \over 8q}}
\eqno(A12)
$$
In the limit $ {Q\over N}\longrightarrow 0$, this reduces to:
$$
Z_0(Q) = {3 \over 16\sqrt{2\pi}}\  {1 \over q^{ 5\over 2}} \
e^{-{1\over 8q}}
\eqno(A13)
$$
\par
We now turn to the calculation of correlation functions.

Consider the generating function for correlation functions, defined by:
$$
G(\vec k(s)) ={1 \over Z_0(Q)}
 \int_{\vec  r(N)=\vec  r(0)}{\cal D}r(s)\ \delta \left({1
\over N} \int^ N_0 {\rm d} s\ \vec  r(s) \right)\delta \left(Q-{1 \over N}
\int^ N_0 {\rm d} s\ \vec  r^2(s) \right)
$$
$$
\times {\rm exp} \left(-{1 \over 2}
\int^ N_0 {\rm d} s\ \vdot  r^2(s) \right) + {i\over N} \int^ N_0 {\rm d} s\
\vec k(s) \ \vec r(s)
\eqno (A14)
$$
The various usual correlation functions can be obtained from (A14), by
taking the proper sum of $\delta$-functions for the function $\vec
k(s)$ ( see below, eq.(A19), (A20) ) .
Defining the Fourier component of $\vec k(s)$ by:
$$
\vec k_n = \int_0^N {\rm d} s \ e^{ i{2\pi \over N} s} \ \vec
k(s)
\eqno(A15)
$$
the generating function can be rewritten as:
$$
G(\vec k^*_n,\vec k_n) = {1 \over Z_0(Q)}
\int_{-\infty}^{+\infty} {{\rm d}z \over 2\pi} \ f(z)
\ \exp\left( -{N\over 2} \sum_{n=1}^{+\infty} { \mid \vec k_n \mid^2 \over
iz + 2 n^2 \pi^2} \right)
\eqno(A16)
$$
where $f(z)$ is given by (A7).
This integral can again be evaluated by the method of residues. The
poles are the same as before, given by (A4), but now,
due to the new exponential term, they are of infinite order.

In the swollen case, $ q \longrightarrow +\infty$ , again only
the pole $n=1$ contributes. The calculation of the residue can be done
and yields:
$$
G(\vec k^*_n,\vec k_n) = e^{-{N \over 4\pi^2} \sum_{n=2}^{+\infty} {
\mid \vec k_n \mid^2 \over n^2 -1 }} \
\sum_{p=0}^{+\infty} {(d-1)! \over p!(p+d-1)!} (-1)^p \left({Q \mid
\vec k_1 \mid^2 \over 2 }\right)^p
\eqno(A17)
$$
which can be written in terms of Bessel functions$^{(15)}$ as:
$$
G(\vec k^*_n,\vec k_n) = (d-1)! \left({2 \over Q \mid \vec k_1 \mid^2}
\right)^{d-1 \over 2} J_{d-1} \left( \sqrt{2 Q \mid \vec k_1 \mid^2}
\right) e^{-{N \over 4\pi^2} \sum_{n=2}^{+\infty} {
\mid \vec k_n \mid^2 \over n^2 -1 }}
\eqno(A18)
$$
The monomer density $\rho(\vec k)$ is obtained by taking:
$$
\vec k(s) = N\ \delta (s-s_0)\ \vec k
\eqno(A19)
$$
and the pair correlation functions used in eq.(19) and (26) are
obtained by taking:
$$
\vec k(s) = N\  (\delta (s-s_0)\ \vec k + \delta (s-s'_0)\ \vec k')
\eqno(A20)
$$
In the large distance limit, $ Q \mid \vec k_1 \mid^2 \longrightarrow 0$,
we obtain :
$$
\rho(\vec k) = N\ e^{- {Q \over 2d} \vec k^2 }
\eqno(A21)
$$
and
$$
G(\vec k^*_n,\vec k_n) = e^{- {Q \over 2d} \mid\vec k_1 \mid ^2
-{N \over 4\pi^2} \sum_{n=2}^{+\infty} {
\mid \vec k_n \mid^2 \over n^2 -1 }}
\eqno(A22)
$$
which is the form used in (19) and (26).

In the collapsed case, $ q \longrightarrow 0$ , all poles contribute, but
in the large distance limit, the calculations simplify, and yield,
for the density, exactly the
same result as above.
\end